\documentstyle[prb,preprint,aps,psfig]{revtex}

\begin{document}

\draft \title{A correlated {\it ab initio} treatment of 
the zinc-blende wurtzite polytypism of SiC and III-V nitrides}

\author{Beate Paulus and Fa-Jian Shi} 
\address{Max-Planck-Institut f\"ur
Physik komplexer Systeme, Bayreuther Stra\ss e 40, 01187 Dresden, Germany}
\author{Hermann Stoll} \address{Institut f\"ur Theoretische Chemie,
Universit\"at Stuttgart, 70550 Stuttgart, Germany}

\maketitle

\begin{abstract}
Ground state properties of SiC, AlN, GaN and InN
in the zinc-blende and wurtzite structures are determined using
an {\it ab initio} scheme.  
For the self-consistent field part of the calculations, the 
Hartree-Fock program {\sc Crystal} has been used. 
Correlation contributions are
evaluated using the coupled-cluster approach with single and
double excitations. This is done by means of increments derived for
localized bond orbitals and for pairs and triples of such bonds.
At the Hartree-Fock level, it turns out that for SiC the 
zinc-blende structure is
more stable although the very small energy difference
to the wurtzite structure is an indication of the experimentally 
observed polytypism.
For the III-V nitrides the wurtzite structure is found to be significantly more
stable than the zinc-blende structure. 
Electron correlations do not change
the Hartree-Fock ground state structures, but energy
differences are enlarged by up to 40\%.
While the Hartree-Fock lattice
parameters agree well with experiment, the
Hartree-Fock cohesive energies reach only 45\% to 70\% of the
experimental values. 
Including electron correlations we
recover for all compounds about 92\% of the experimental 
cohesive energies.
\end{abstract}

\section{Introduction}

SiC and the III-V nitrides (especially GaN) are the most
promising wide band-gap semiconductors for short-wavelength 
optoelectronics and for high-power, high-temperature microelectronic
devices \cite{davis93}. But it is difficult to grow high-quality single
crystals of these materials. This is due to the polytypism occurring between
zinc-blende (ABCABC... sequence along the 111 direction, see 
Fig.\ref{structure}a)
and wurtzite structure (ABABAB... sequence along the 100 direction,
see Fig.\ref{structure}b).
SiC shows many more complicated stacking sequences built up as
a mixture between the fully cubic structure (zinc-blende) and the fully
hexagonal structure (wurtzite) with different percentages of
hexagonality. Therefore it is important to 
understand the bulk properties of these materials, from a theoretical point 
of view, at least for the
two simplest structures (zinc-blende and wurtzite).\\
Over the past decade, {\it ab initio} calculations based on density
functional theory with a local-density approximation (LDA) have
been performed for these materials. In some of them, 
the energy differences between the zinc-blende and the
wurtzite structure have been evaluated
\cite{yeh92,min92,miwa93,kim96,karch94,kaeckell94}. 
There is no 
possibility to measure these energy differences, thus a complement by -- and
a comparison with -- other {\it ab initio} methods is desirable. 
{\it Ab initio} Hartree-Fock self consistent field (HF) calculations
for solids\cite{pisani88} are feasible nowadays, using the program package
{\sc Crystal}\cite{crystal92}. They have the merit
of treating the non-local exchange exactly although lacking electron 
correlations 
per definition. In LDA calculations where both exchange and 
correlations are covered in
an implicit way,
a systematic improvement towards the exact
results appears to be
difficult. Treating the non-local exchange exactly yields
a better (microscopic) understanding of the electron interaction in these
materials and, at the same time, leads to a good starting point for a
post-Hartree-Fock correlation treatment.\\
Electron correlations can be taken into account explicitly using 
many-body wave-functions of the
configuration-interaction or coupled-cluster type.
These methods are well developed for finite systems like
atoms and molecules. Infinite systems such as solids require a
size-consistent approach which is achieved at the 
coupled-cluster level. Because of the local
character of the correlation hole
one can expand the correlation energy of the solid in terms of local increments
\cite{stoll92}.  The idea thereby is to determine the required matrix
elements by studying local excitations in finite clusters which are
accessible to a full post-HF quantum-chemical treatment.
We apply this idea to the zinc-blende and
the wurtzite structures of the title materials.\\
In the first part of this paper  we want to report on
Hartree-Fock calculations for SiC, AlN, GaN and InN using 
pseudopotentials and 
optimized gaussian basis sets for the zinc-blende and wurtzite structure
(Sec. II). 
The method of increments, which is briefly
described in Sec. IIIa, is extended to the wurtzite structure;
computational details are given in Sec. IIIb.
We discuss results for the energy differences, the lattice parameters,
the cohesive energies and the bulk moduli in
Section IV. Conclusions follow in Section V. 

\section{Hartree-Fock calculation}

The periodic Hartree-Fock method as implemented in
the {\sc Crystal92} program \cite{crystal92} is used in our calculations.
The problem of accurately calculating the Coulomb and exchange 
contributions to the Fock operator is addressed by taking 
very tight tolerances 
in the evaluation of these series, which lead to convergence of the total 
energy to about
$10^{-4}$ Hartree. For the inert core electrons we use the 
scalar-relativistic
energy-consistent pseudopotentials of Bergner {\it et al.}\cite{bergner93}.
Two exceptions are made: 3-valence-electron pseudopotentials for the 
post-$d$ elements Gallium
and Indium have been found to underestimate the closed-shell
repulsion of the underlying $d$ shell on valence electrons of
neighbouring atoms\cite{schwerdt95}; we therefore performed
the SCF calculations for Ga and In compounds with 13-valence electron
pseudopotentials\cite{13vepp}, explicitly treating the highest
occupied $d$ shell. \\
For all of these pseudopotentials corresponding
atomic basis sets have been optimized \cite{bergner93,13vepp}. These are used
in the Hartree-Fock calculations for the free atoms, which are
performed with the program package {\sc Molpro94}
\cite{molpro94}.
For the solid we generated contracted [$2s2p1d$] gaussian valence
basis sets (Ga and In:  [$2s2p2d$])
as follows:
starting from the energy-optimized atomic basis sets
just mentioned, the inner functions
of $s$ and $p$ symmetry are contracted using atomic
ground-state orbital coefficients; the outermost $s$ and $p$ functions
are left uncontracted. We re-optimized the exponents of these 
functions, together with an additional $d$
polarization function, for the solid. Thereby, we must pay attention 
to the fact, that too diffuse exponents cause numerical problems in 
{\sc Crystal}. But the most diffuse exponents of the atom,
which are necessary for a correct description of the free atom, have little
effect in the solid, since due to close-packing basis functions of
the neighbouring atoms take over their role. 
The crystal-optimized basis sets are listed in Table \ref{basis}.\\
In order to check the quality of the basis sets we calculated the ground-state
energy of the zinc-blende structure with  larger [$3s3p1d$]
and [$3s3p2d$] basis sets for Si/C/Al/N and Ga/In, respectively. These 
basis sets are generated 
from the latter ones by de-contracting the outermost primitive gaussian of 
the inner $s$ and $p$ functions, thus
leaving two outer gaussians uncontracted. The $d$ functions are kept unchanged.
There are very small changes in the cohesive energy due to this enlargement of
the basis. The maximum deviation occurs for InN with 0.005 a.u..
An additional test of the basis set chosen is the comparison with the
Hartree-Fock values by Pandey {\it et al.}\cite{pandey96} who calculated the
ground state of GaN in the wurtzite structure with an all-electron basis
of comparable quality: the deviation from our values is less than
0.4\%  for the lattice constant.\\
The geometry optimization is performed as follows: For the 
zinc-blende structure there is only one free parameter, the cubic lattice
constant $a_{\rm zb}$. This is varied in steps of 1\% of the
experimental value. Six points are calculated and a 
quadratic fit is applied to determine the position
of the minimum and the curvature.\\
The wurtzite structure has three free geometry parameters.
In the first place, the two lattice constants $a_{\rm w}$ 
(in the hexagonal plane)
and $c_{\rm w}$ (perpendicular to the hexagonal plane), or one lattice
constant and the ratio $\frac{c}{a}$ which in the ideal case 
(bond lengths as in the zinc-blende structure) 
is $\sqrt{\frac{8}{3}}= 1.6330$, have to be optimized.
The third parameter is the cell-internal dimensionless constant $u$, 
which denotes the
position of the second atom along the $c$-axis. The ideal value would be
$\frac{3}{8}$; a deviation from it corresponds to a change in the bond angle
away from the ideal tetrahedral one. We optimize all these
parameters in the following way: First we vary the volume of the 
unit cell in steps of 1\% and determine $V_{\rm min}$, then, 
for fixed $V_{\rm min}$, we vary the ratio $\frac{c}{a}$ and last, for
fixed $V_{\rm min}$ and $\left(\frac{c}{a}\right)_{\rm min}$,
the cell-internal $u$. With the optimized $u$, we checked its influence on 
$\frac{c}{a}$ and $V$, which in general is very small.
 
\section{Correlation calculations}
\subsection{Methods of increments}

Here we only want to sketch the basic ideas and some important formulae
of the method of increments. A formal derivation and more details of
the method for an infinite periodic system can be found in Ref.\
[\onlinecite{paulus95}].
The method relies
on localized bond orbitals generated in a SCF reference calculation.
One-bond correlation-energy increments $\epsilon_i$ are obtained by
correlating each of the localized orbitals separately while keeping
the other ones inactive. In the present work we are using the
coupled-cluster approach with single and double substitutions
(CCSD).  This yields a first approximation of the correlation energy
\begin{equation} E_{\mbox{corr}}^{(1)}=\sum_i \epsilon_i ,
\end{equation} which corresponds to the correlation energy of
independent bonds.\\ In the next step we include the correlations of
pairs of bonds. Only the non-additive part $\Delta\epsilon_{ij}$ of
the two-bond correlation energy $\epsilon_{ij}$ is needed.
\begin{equation}
\Delta\epsilon_{ij}=\epsilon_{ij}-(\epsilon_i+\epsilon_j).
\end{equation} Higher order increments are defined analogously. For
the three-bond increment, for example, one has \begin{equation} \Delta
\epsilon_{ijk}= \epsilon_{ijk} -( \epsilon_i +  \epsilon_j +
\epsilon_k) -(\Delta \epsilon_{ij}+\Delta \epsilon_{jk}+ \Delta
\epsilon_{ik}).  \end{equation} The correlation energy of the solid
is finally obtained by adding up all the increments with appropriate
weight factors:  \begin{equation} E_{\mbox{corr}}^{\mbox{solid}} =
\sum_i  \epsilon_i + \frac{1}{2}\sum_{ij \atop i \not= j} \Delta
\epsilon_{ij}+ \frac{1}{6}\sum_{ijk \atop i \not= j \not= k} \Delta
\epsilon_{ijk} + ... .  \end{equation} It is obvious that by
calculating higher and higher increments the exact correlation energy
within CCSD is determined. However, the procedure described above is only
useful if the incremental expansion is well convergent, i.e. if
increments up to, say, three-bond increments are sufficient, and if increments
rapidly decrease with increasing distance between localized
orbitals.
These conditions were shown to be well met in the case of
elementary semiconductors \cite{paulus95} and cubic III-V semiconductors
\cite{paulus96}.

\subsection{Computational details}

\begin{figure}
\psfig{figure=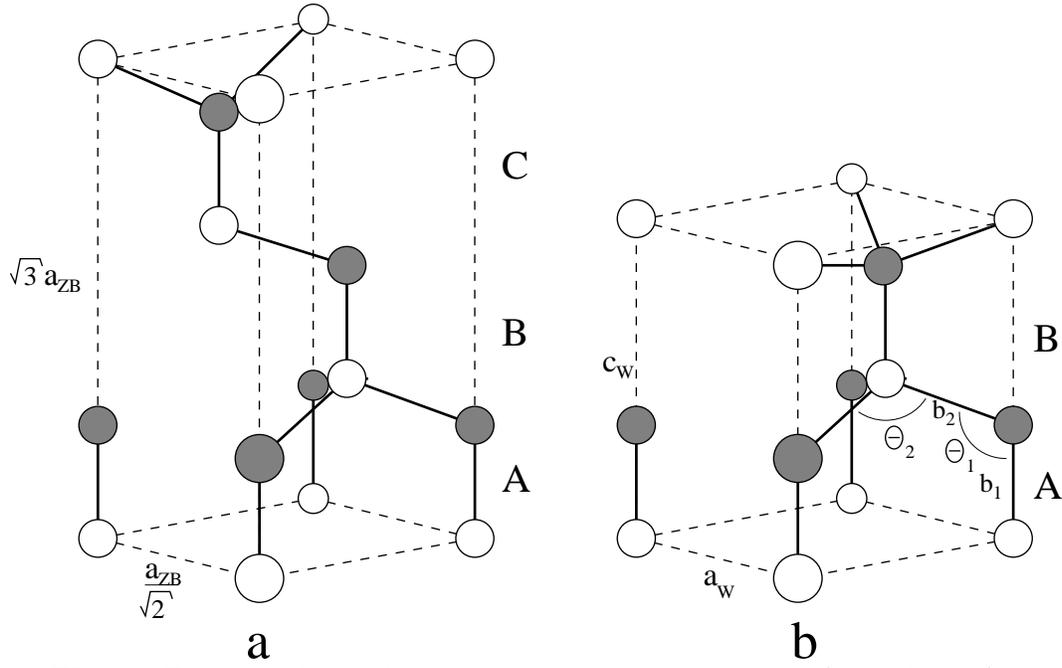,width=14cm,angle=-90}
\caption{\label{structure}Figure a shows the primitive 
hexagonal unit cell (dashed lines)
of the zinc-blende structure, figure b the one of the wurtzite structure.
The solid lines mark the \protect{$sp^3$} bonds. The stacking sequences
ABC (zinc-blende) and AB (wurtzite) are indicated, too. Open and shadded 
circles refer to the two different atoms of the structures.} 
\end{figure}

\begin{figure}
\centerline{\psfig{figure=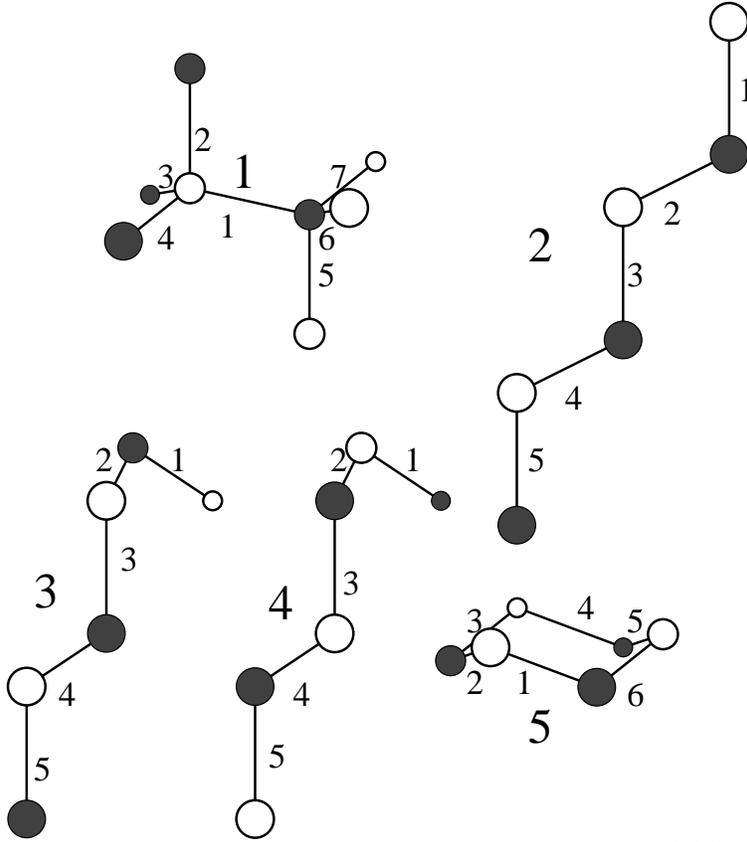,width=10cm,angle=-90}}
\caption{\label{zb}X$_n$Y$_n$H$_m$-clusters of the zinc-blende structure
treated at the  CCSD level for the correlation calculation;
big numbers designate clusters,
small numbers the bonds in each cluster;
H-atoms are not drawn.}
\end{figure}

\begin{figure}
\centerline{\psfig{figure=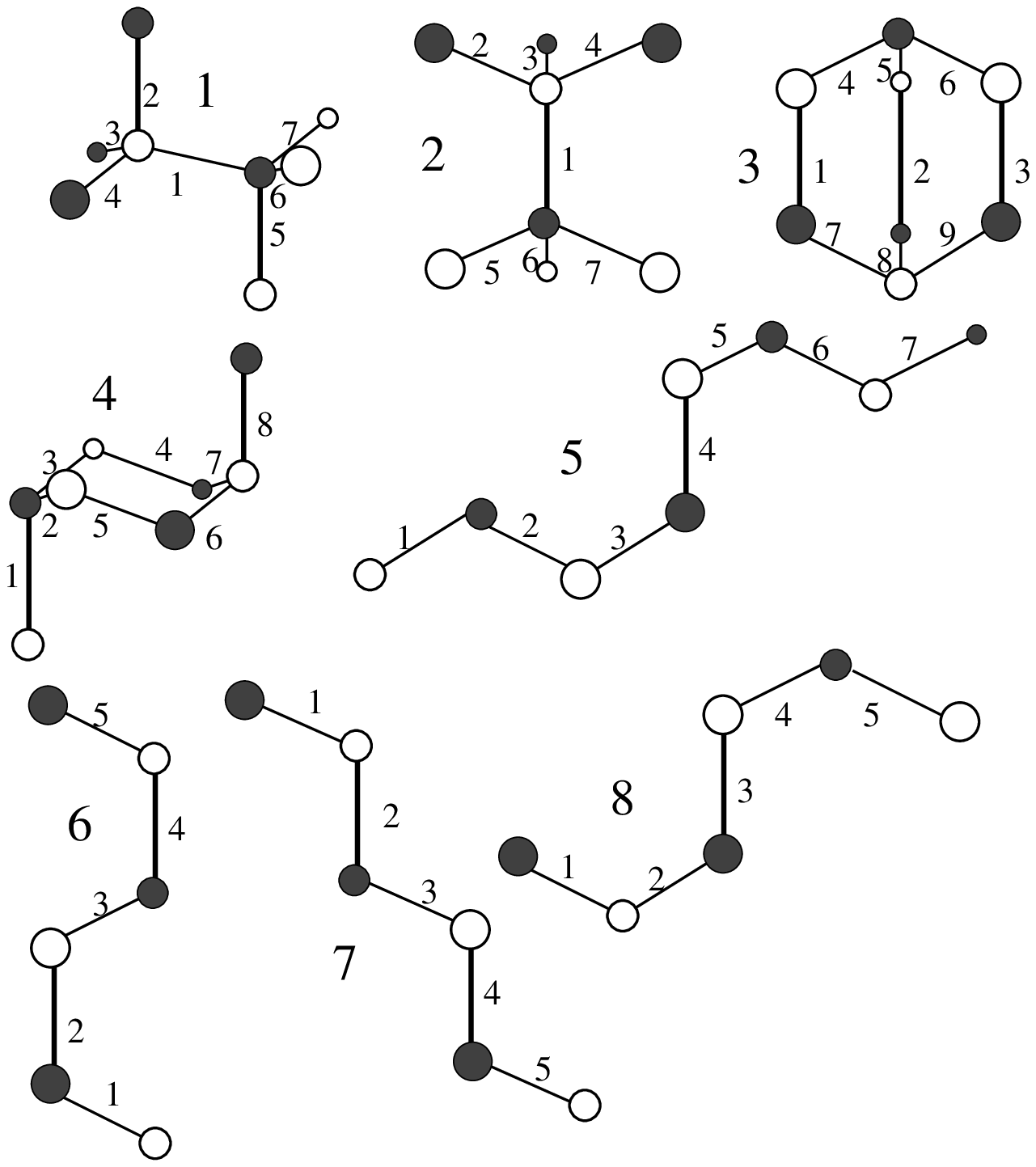,width=10cm,angle=-90}}
\caption{\label{wurt}X$_n$Y$_n$H$_m$-clusters of the wurtzite structure treated \
at
the CCSD level for the correlation calculation; the broader lines indicate
the vertical bonds;
big numbers designate clusters,
small numbers the bonds in each cluster;
H-atoms are not drawn.}
\end{figure}

We apply the  procedure described above to calculate the correlation energies per unit cell (u.c.)
for SiC and the III-V
compounds AlN, GaN and InN, with the zinc-blende (2 atoms/u.c.) and wurtzite (4 atoms/u.c.)
structures.
We evaluate the symmetry-unique increments in Eq.\ (4) and multiply them by appropriate weight factors 
which are determined by the  crystalline symmetry of the unit cell.
For the zinc-blende structure
all four bonds of the unit cell are equivalent.
In the wurtzite structure the two vertical bonds
along  the c-direction ($b_1$ in Fig.\ref{structure}b) differ from  
the six bonds
of the buckling plane ($b_2$ in Fig.\ref{structure}b); 
these bonds will be called 
planar bonds in the following.
For a direct comparison of the two structures we reduce the
correlation energy per unit cell of the wurtzite structure by a factor 2.
The weight factors of
all increments considered for both structures are listed in Table
\ref{inc}.\\
Since (dynamical) correlations are a local effect, the increments should be
fairly local entities at least for semiconductors and insulators. We
use this property to calculate the correlation energy increments in finite
clusters. We select the clusters as fragments of the zinc-blende and
wurtzite structures so that we can calculate all two-bond increments
up to third-nearest neighbours and all nearest-neighbour three-bond increments.
The clusters used for the zinc-blende structure are shown in Fig.\ref{zb},
those for the wurtzite structure in Fig.\ref{wurt}.
In the zinc-blende structure all bond
lengths and bond angles (tetrahedral, $\Theta$=109.4712$^{\circ}$)
are the same. As mentioned before, in the  wurtzite structure
the vertical and planar bond lengths are different, and so are the
planar-vertical ($\Theta_1$ in Fig.\ref{structure}b)  and planar-planar
($\Theta_2$ in Fig.\ref{structure}b) bond angles, too.
The numerical values for the bond lengths and bond angles are taken from the Hartree-Fock calculations for
the periodic solid; for the zinc-blende
structure these are $b^{\rm SiC}$=1.8963\AA,
$b^{\rm AlN}$=1.8941\AA, $b^{\rm GaN}$=1.9579\AA\
and $b^{\rm InN}$=2.1594\AA, for the wurtzite structure we have
$b_1^{\rm SiC}$=1.9089\AA, $b_2^{\rm SiC}$=1.8919\AA,
$\Theta_1^{\rm SiC}=109.3706^{\circ}$, $\Theta_2^{\rm SiC}=109.5717^{\circ}$;
$b_1^{\rm AlN}$=1.8983\AA, $b_2^{\rm AlN}$=1.8934\AA,
$\Theta_1^{\rm AlN}=108.3516^{\circ}$, $\Theta_2^{\rm AlN}=110.5684^{\circ}$;
$b_1^{\rm GaN}$=1.9619\AA, $b_2^{\rm GaN}$=1.9547\AA,
$\Theta_1^{\rm GaN}=109.0063^{\circ}$, $\Theta_2^{\rm GaN}=109.9321^{\circ}$;
$b_1^{\rm InN}$=2.1677\AA, $b_2^{\rm InN}$=2.1608\AA,
$\Theta_1^{\rm InN}=108.8072^{\circ}$, $\Theta_2^{\rm InN}=110.1270^{\circ}$.
The dangling bonds of the clusters are saturated by hydrogens.
The internuclear X---H distances have been taken from the ${\rm XH}_4$ molecule for C and Si
($b_{\rm CH}$=1.102\AA\ and $b_{\rm SiH}$=1.480\AA), for the other cases
they have been optimized in CCSD calculations for ${\rm XNH}_6$ cluster,
yielding $b_{\rm AlH}$=1.614\AA, $b_{\rm GaH}$=1.621\AA,
$b_{\rm InH}$=1.711\AA\, and for ${\rm AlNH}_6$ yielding $b_{\rm NH}$=1.016\AA.
Using the finite clusters instead of the periodic solid,
we can use the Foster-Boys criterion \cite{foster60} to localize 
bond orbitals instead of constructing Wannier functions.
Following the procedure described above we calculated
the correlation-energy increments at the CCSD  level, using the program
package MOLPRO94 \cite{molpro94}, successively correlating more and more of
the localized X---Y bond orbitals while keeping the other cluster orbitals inactive.\\
The one-particle basis sets used in the cluster calculations can be characterized as follows.
For hydrogen we chose Dunning's double-$\zeta$ basis\cite{dunning89} 
without the $p$ polarization
function. For the other elements we use energy-consistent quasi-relativistic
large-core pseudopotentials\cite{bergner93} with the corresponding valence
basis sets optimized for the atoms. One basis set (basis A) has the same
quality as that for
the corresponding Hartree-Fock calculations of the solid:
$(4s4p)/[3s3p]$ supplemented
with one $d$ polarization function, which is optimized in a CCSD calculation
for the ${\rm XYH}_6$ cluster\cite{paulus95,paulus96}. In addition, an extended basis set
(basis B) has been generated by uncontracting
the $s$ and $p$ functions of basis A and by replacing the single
$d$ function by a $2d1f$ polarization set\cite{paulus95,paulus96}.
This basis set is only used for the five largest increments, which are
evalueated in ${\rm X}_2{\rm Y}_2{\rm H}_{10}$ clusters. \\
The convergence of the incremental expansion has been checked for all
substances, but will be discussed here in detail only for SiC.
In Table \ref{inc} all required increments are listed for both structures.
They rapidly decrease with increasing distance of the bonds involved. 
The nearest-neighbour two-bond
increments are about eight times larger than the weighted sum of all 
third-nearest ones.
After the latter, we truncate
the expansion of the correlation energy because with the 
fourth-nearest neighbours'
contribution we reach the van der Waals limit of the correlation energy
(decay $\sim r^{-6}$). The error due to the truncation can be 
estimated to 0.5\%
of the correlation energy.
The convergence with respect to the number of bonds simultaneously correlated
in the incremental expansion is also satisfactory.
For SiC, the one-bond increment contributes with 38\% of the correlation
energy, the two-bond increments are very important (67\%) but lead to an overestimation of
the correlation energy, which is effectively reduced by the three-bond increments (5\%).
The most important four-bond increments contribute only with 0.5\%
to the incremental expansion, and are neglected therefore.
Thus, the overall error due to truncation of the incremental
expansion is less than 1\% of the correlation energy.
The shortcomings of one- and many-particle basis sets in the
determination of individual increments cause the largest part of
the error of the correlation energy. The dependence on the one-particle basis
is discussed in detail in Ref.\ [\onlinecite{paulus96}] for the III-V
compounds.

\section{Results and Discussion}

\subsection{Energy difference between wurtzite and
zinc-blende structure}

As described in section II,  we separately optimized the 
geometries of the 
zinc-blende and wurtzite structures and determined the total 
Hartree-Fock energy for each structure at its equilibrium geometry.
We compare the two total energies and calculate their
difference, $\Delta E_{\rm w-zb}^{\rm HF}$, per two atoms in a.u..
In the second step, we calculated correlation energies,
at the Hartree-Fock optimized
geometries, for both the zinc-blende and the wurtzite structure
using the method of increments described above. The energy difference
between the two structures $\Delta E_{\rm w-zb}^{\rm corr}$ is calculated for
the correlation part.
The results are listed in the first column of Table \ref{results}.
In all cases, electron correlation yields the same energetic order of states as Hartree-Fock,
but energy differences are enlarged by
up to 40\% for SiC and for none of the substances are negligible.\\
For SiC we found that the zinc-blende structure is slightly
more stable than the wurtzite one, but the energy difference is
so small, that it might be taken as an indication of the occurring polytypism. 
The LDA calculations lead to the same conclusion, but the
absolute value of the relaxed LDA energy difference \cite{kaeckell94} 
is even smaller than ours, by about a factor of 10. K\"ackell {\it et al.}
found a strong influence of the relaxation on the energy difference
(up to a factor of 10) which explains the large discrepancy
between the two LDA results listed in Table III.
The large deviation between our result and
the relaxed LDA result can be understood from a 
comparison of the corresponding volumes
of the unit cell. Whereas in the work of K\"ackell {\it et al.}
the zinc-blende volume was found to be smaller by about 0.18 \AA$^3$ than 
the wurtzite one,
in our calculations the volumes are essentially the same. 
\\
For all III-V nitrides the calculations yield
a stable wurtzite structure as experimentally observed. The energy differences
are significantly larger than the one for SiC, which clearly indicates that 
no polytypism exists for the III-V nitrides. The most stable compound 
in wurtzite structure is AlN, followed by InN and then GaN.
For the latter two systems it was also possible to epitaxially grow 
the zinc-blende structure \cite{lei92,strite92}. In comparison 
with the LDA results by Yeh {\it et al.}\cite{yeh92}, our energy differences 
are larger again, by about a 
factor of two, but the trends 
reproduced are the same.
A possible explanation is the following:
The LDA gets the long-range interaction right, because the latter can be well
described in a mean-field approach, and thus yields the right
structure. But the short-range correlations, which can only be described with a many-body theory,
have a substantial influence on the
total energy and on the energy differences as well.

\subsection{Lattice parameter}

In order to find the minimum of the total energy we calculated the optimized 
Hartree-Fock geo\-metries for both the zinc-blende and wurtzite structures.
The results are listed in Table \ref{results}. The experimental 
values\cite{landoltiii17a} are measured 
at room temperature. The extrapolation to zero Kelvin 
changes the measured value for SiC by 0.001\AA. 
In Table \ref{results} we listed for comparison the range of the
room-temperature experimental values 
where available and some other
theoretical values obtained within the LDA.\\
For SiC the lattice constants of both
structures are slightly too large (by 0.5\%), but the $\frac{c}{a}$-ratio
of the wurtzite structure perfectly agrees with experiment.
This ratio is somewhat larger than the ideal one, which indicates
that the zinc-blende structure is more stable than the wurtzite one.
The internal-cell parameter $u$ only marginally differs from the 
ideal one. The LDA results underestimate the lattice constants
(Karch {\it et al.}: 0.2\% ... 0.3\%;K\"ackell {\it et al.}: 0.7\% ... 1.8\%). 
Their $\frac{c}{a}$-ratio is in good agreement
with experiment, too. Karch {\it et al.} did not allow for an internal cell
relaxation and K\"ackell {\it et al.} found it even closer to the 
ideal one than we did.\\
The III-V nitrides are stable in the wurtzite structure. The
$\frac{c}{a}$-ratio is smaller than the ideal one, i.e. the bonds in the 
direction of the $c$-axis (Fig.1b, $b_1$) are shorter than
those ((Fig.1b, $b_2$) of the buckling plane. \\
Our calculated lattice parameters for the wurtzite structure of AlN agree well 
with experiment, the $\frac{c}{a}$-ratio slightly differs from experiment
and a larger deviation from the ideal $u$ occurs than for SiC. For the zinc-blende
structure there do not exist any experimental data, but our result
lies in the same region as the various LDA results.\\
For GaN the Hartree-Fock calculation overestimates the lattice
constants by $\sim$1.3\%, although the $\frac{c}{a}$-ratio agrees
well with experiment. This fact can partly be explained with the missing
electron correlation (cf.\ the discussion below).
 The $u$-parameter is smaller
than that of AlN but still larger than the ideal one, which indicates
that the wurtzite structure of GaN is not as stable as that
of AlN.\\
For InN all lattice parameters are close to experiment. 
The $u$-parameter lies between those of AlN
and GaN, the same holds true for the $\frac{c}{a}$-ratio.\\
The influence of correlations on the lattice parameters has been discussed in
detail for various cubic III-V compounds, applying the method of increments, by
Kalvoda {\it et al.}\cite{kalvoda97}. Here, we checked this influence for
SiC only where we calculated the cubic lattice constant at the correlated level
with basis A. Virtually no change to the Hartree-Fock value is seen
(an increase $\leq$ 0.1\%). This can be understood by the opposing effect of
inter- and intra-atomic correlations.
The inter-atomic correlation that dominates for the lighter elements such as C and N
increases the lattice constant, while intra-atomic correlations that become
more important for heavier elements decrease it. Thus the net effect of correlation cancels
for SiC and probably, to a great part, for AlN, too. GaN is an exception:
there the core-valence correlation yields an important 
contribution to the  reduction
of the lattice constant, because the fully occupied 3$d$ shell of Ga is 
easily polarizable. Simulating this effect with a core polarization
potential reduces the lattice constant by about 1\% (e.g. for
GaAs by 0.9\%\cite{kalvoda97}).  \\
The LDA results agree on average quite well with experiment, but are not
 systematically better than our results.
With LDA,
there is a small underestimation of the lattice constants, which 
is a well-known shortcoming of this method.
Moreover, the LDA results spread quite a lot.
       
\subsection{Cohesive energies}

The cohesive energy per unit cell is calculated as difference
between the total energy of the zinc-blende structure and the free atoms:
$E_{\rm coh}=E_{\rm zb}-\sum_i E_{{\rm atom},i}$. For the 
Hartree-Fock calculation, we use the enlarged 
[$3s3p1d$] (Ga und In: [$3s3p2d$]) basis, although the changes
to the [$2s2p1d$] basis are small. 
The Hartree-Fock results together with
the experimental data are listed in Table \ref{coh}.\\
The energy differences between the zinc-blende and wurtzite structure 
are so small, that they could not be measured so far. We take the experimental 
cohesive energies from Harrison \cite{harrison89} and correct them 
by the phonon
zero-point energies $\frac{9}{8}k_B\Theta_D$ (derived from the Debye
model \cite{farid91}; $\Theta_D$ from Ref. \onlinecite{landoltiii17a})
as well as by atomic spin orbit splittings \cite{moore}.
Both effects increase the cohesive energy. For SiC and AlN the phonon
zero-point energy dominates with about 0.004 a.u. and 0.003 a.u.,
respectively, over the spin-orbit splitting of the free atoms
(by about a factor of 10). For GaN both effects are nearly equal 
($\sim$0.002 a.u.), for InN the spin-orbit splitting clearly dominates
($\sim$0.007 a.u.).
In all cases, the total effect is larger than the energy difference 
between the two structures.
According to Harrison \cite{harrison89} an experimental error d
ue to measuring the heat of formation
and the heat of atomization  at different temperatures can be estimated
to about 0.0015 a.u. which leads to an experimental error bar of less than
1\% of the cohesive energy.\\
At the Hartree-Fock level, we reach  
between 70\% (SiC) and 45\% (InN) of the experimental cohesive energy. 
At the correlated level, we perfomed calculations with two different
basis sets. With basis A we reach 89\% of the experimental cohesive
energy for SiC and AlN,  and 86\% and 81\% for GaN and InN, respectively
(Table \ref{coh}).
The improved basis set (basis B) yields
much better results especially for the heavier compounds:
whereas the improvement due to the basis set is only 4\% for SiC and AlN,
for GaN and InN it is up to 7\% (Table \ref{coh}).
This shows, that
excitations
from the $sp^3$ bonds into unoccupied $d$ and $f$ shells are becoming
more and more important. Overall we reach on average 91\% of the
experimental cohesive energy.
For comparison LDA results are listed in Table \ref{coh}, too. They
clearly overestimate the cohesive energies (by up to 44\%). 

\subsection{Bulk moduli}

For cubic structures the bulk modulus 
$B=V\frac{\partial^2 E}{\partial V^2}$ can be easily derived from the
curvature:
\begin{equation}
\label{eqbulk}
B = \left( \frac{4}{9a}\frac{\partial^2 }{\partial a^2}
- \frac{8}{9a^2}\frac{\partial }{\partial a} \right) E(a).
\end{equation} 
Calculation of the bulk modulus for the wurtzite structure would require
full geometry optimization at every volume point. Only thus,
the correct relaxation behaviour to a homogeneous pressure would be obtained.
We have not performed this time-demanding procedure and restrict ourselves to
the bulk moduli of the zinc-blende structure here. The latter 
have been determined
at the optimized Hartree-Fock lattice constant, so that the second term 
in (\ref{eqbulk}) is zero. The results are listed in Table \ref{bulk}.\\
The Hartree-Fock approximation leads to an overestimation of the 
bulk moduli by between 10\%
and 27\% .  Electron correlations 
reduce the bulk moduli by allowing an instantaneous 
response of the electrons to an homogeneous pressure.
For a detailed analysis and a discussion of the errors 
see Ref. \onlinecite{kalvoda97}. Here we performed correlated calculations
for zinc-blende SiC in basis A only, where electron 
correlations reduce the bulk 
modulus only slightly (see Table \ref{bulk}). A further improvement 
of the one-particle
basis set is expected to yield a further reduction. 
The LDA results are in better agreement with experiment than ours. 
In Table \ref{bulk}, we list
LDA results for both zinc-blende and wurtzite structure in order  
to show that the structural influence on the bulk moduli is small.
Kim {\it et al.}\cite{kim96} even claim that the uncertainty of 
their calculations is larger 
than the difference between the bulk moduli of the two structures.

\section{Conclusion}

We have performed correlated {\it ab initio} calculations for SiC and III-V
nitrides in the zinc-blende and wurtzite structures. The mean-field part
has been determined for the periodic solid using the program package 
{\sc Crystal92}.  Correlation contributions have been
evaluated using the coupled-cluster approach with single and
double excitations, applying the methods of local increments.
Pseudopotentials have been used in conjunction with 
valence basis sets optimized for the solid.
Results have been obtained for the ground-state energies of both 
structures with fully
optimized geometries. For SiC the zinc-blende structure is slightly more 
stable, but
the small energy difference to the wurtzite structure confirms the occurring
polytypism. AlN, GaN and InN are stable in the wurtzite structure with a 
significant energy difference to the zinc-blende structure. The 
calculated lattice 
parameters agree well with experiment. The cohesive energies reach 
only 
45\% to 70\% of the experimental values, at the Hartree-Fock level, but electron correlations
increase these percentages to over 90\%.
Bulk moduli for the zinc-blende structure are overestimated by 10\% to 30\%
at the Hartree-Fock level, electron correlations yield a reduction.

\section{Acknowledgements}

We gratefully acknowledge the valuable scientific discussions and
suggestions of Prof.\ P. Fulde, Dresden.
We are grateful to 
Prof.\ R.\ Dovesi, Torino, and Prof.\ H.-J.\ Werner, Stuttgart, 
for providing their programs {\sc Crystal} and {\sc Molpro} respectively.
We also thank Dr.\ T.\ Leininger,
Stuttgart,  and Dr.\ M.\ Dolg, Dresden, for providing the Ga and In
small-core pseudopotentials prior to publication.

\renewcommand{\baselinestretch}{1.2}
\begin{table}
\caption{\label{basis}Crystal-optimized basis sets for
SiC, AlN, GaN and InN.}
\begin{tabular}{c||c|c||c|c||c|c}
&$s$-exp.&coeff.&$p$-exp.&coeff.&$d$-exp.&coeff.\\
\hline
C&2.263101&0.496548&8.383025&-0.038544&0.55&1\\
&1.773186&-0.422391&1.993132&-0.203185&&\\
&0.408619&-0.599356&0.559543&-0.498176&&\\
&0.159175&1&0.156126&1&&\\
\hline
Si&4.014378&-0.039508&1.102481&0.084583&0.40&1\\
&1.393707&0.296150&0.583127&-0.185748&&\\
&0.251658&-0.599752&0.208675&-0.554852&&\\
&0.135&1&0.15&1&&\\
\hline
Al&2.786337&-0.046411&0.983794&0.052036&0.29&1\\
&1.143635&0.274472&0.358245&-0.155094&&\\
&0.170027&1&0.15&1&&\\
\hline
Ga&32.95500&0.000215&2.562424&0.017921&76.20527&0.007822\\
&8.306842&0.007419&1.450154&-0.100112&25.52835&0.068978\\
&1.349536&0.063868&0.396817&0.109562&9.465050&0.202734\\
&1.145804&-0.348833&0.17&1&3.882911&0.401034\\
&0.294700&0.212388&&&1.504741&0.410996\\
&0.15&1&&&0.502918&1\\
\hline
In&1.744487&0.279421&1.606834&0.130961&16.741457&0.012997\\
&1.055194&-0.628094&1.168418&-0.247565&4.550192&0.187048\\
&0.176720&0.539402&0.200692&0.326411&1.815414&0.442393\\
&0.12&1&0.12&1&0.725833&0.427724\\
&&&&&0.263986&1\\
\hline
N&32.656839&-0.013794&12.146974&-0.041296&0.82$^a$&1\\
&4.589189&0.129129&2.884265&-0.214009&&\\
&0.706251&-0.568094&0.808564&-0.502783&&\\
&0.216399&1&0.222163&1&&\\
\hline
\multicolumn{7}{l}{\footnotesize $^a$ 0.82 is the $d$ exponent optimized
for AlN, for GaN and InN it is 0.75.}
\end{tabular}
\end{table}

\begin{table}
\caption{\label{inc}Correlation-energy increments for
SiC (in a.u.) for the zinc-blende and wurtzite structures,
determined at the CCSD level using basis A. For
the numbering of the clusters and bonds involved, see 
Figs.\protect{\ref{zb}} and \protect{\ref{wurt}} . }
\begin{tabular}{|c|c|c|c|c|c|c|}
& \multicolumn{3}{c}{zinc-blende structure}&\multicolumn{3}{c}
{wurtzite structure}\\
\hline
&cl./bond&increment&weight&cl./bond&increment&weight\\
\hline\hline
$\Delta \epsilon_{i}$&1/1&-0.022471&4&1/1&-0.022428&3\\
&&&&2/1&-0.022511&1\\
\hline
$\Delta \epsilon_{ij}$&1/1,2&-0.016122&6&1/1,4&-0.016190&3\\
&1/1,5&-0.003929&6&1/1,6&-0.003901&3\\
&&&&2/1,2&-0.016076&3\\
&&&&2/1,5&-0.003932&3\\
\hline
$\Delta \epsilon_{ij}$&1/2,5&-0.000773&12&1/3,6&-0.000773&6\\
&1/2,6&-0.000682&24&1/3,7&-0.000681&6\\
&&&&2/2,6&-0.000819&6\\
&&&&2/2,5&-0.000640&3\\
&&&&1/2,5&-0.000774&3\\
&&&&1/2,6&-0.000680&6\\
&&&&1/3,5&-0.000690&6\\
\hline
$\Delta \epsilon_{ij}$
&2/1,4&-0.000247&6&3/1,2&-0.000265&3\\
&2/2,5&-0.000089&6&3/6,7&-0.000152&6\\
&5/1,4&-0.000249&12&4/3,6&-0.000271&3\\
&3/1,4&-0.000117&12&4/1,6&-0.000131&6\\
&3/2,5&-0.000195&24&4/2,8&-0.000205&6\\
&4/1,4&-0.000161&12&5/1,4&-0.000200&6\\
&4/2,5&-0.000076&24&5/4,7&-0.000078&6\\
&&&&5/2,5&-0.000088&6\\
&&&&5/3,6&-0.000167&6\\
&&&&6/2,5&-0.000287&3\\
&&&&6/1,4&-0.000098&3\\
&&&&7/1,4&-0.000187&6\\
&&&&7/2,5&-0.000070&6\\
&&&&8/1,4&-0.000165&6\\
&&&&8/2,5&-0.000076&6\\
&&&&2$^{\rm zb}$/1,4&-0.000248&3\\
&&&&2$^{\rm zb}$/2,5&-0.000088&3\\
&&&&3$^{\rm zb}$/2,5&-0.000196&6\\
&&&&4$^{\rm zb}$/2,5&-0.000076&6\\
\hline
$\Delta \epsilon_{ijk}$&1/1,2,3&0.002013&4&1/1,3,4&0.001998&1\\
&1/1,5,6&0.000138&4&1/1,6,7&0.000130&1\\
&1/1,2,5&0.000040&12&2/1,2,3&0.002024&3\\
&1/1,2,6&0.000134&24&2/1,5,6&0.000139&3\\
&&&&2/1,2,7&0.000057&3\\
&&&&2/1,2,5&0.000181&6\\
&&&&1/1,2,5&0.000040&3\\
&&&&1/1,4,6&0.000133&6\\
&&&&1/1,4,7&0.000038&6\\
&&&&1/1,2,6&0.000132&6\\
&&&&1/1,3,5&0.000131&6
\end{tabular}
\end{table}

\begin{table}
\caption{\label{results}Energy differences and lattice parameters 
of the zinc-blende and wurtzite structure
of SiC and the III-V nitrides. \protect{$\Delta E_{\rm w-zb}$}
is the relaxed energy difference in Hartree per two atoms.
\protect{$a_{\rm zb}$} (in \AA) is the zinc-blende lattice constant;
\protect{$a_{\rm w}$} and \protect{$c_{\rm w}$} are the lattice
constants of the wurtzite structure and $u$ is its cell-internal 
parameter. For comparison other theoretical values and the experimental values 
at room temperature are 
given.}
\begin{tabular}{cc||c||c||c|c|c|c}
&&$\Delta E_{\rm w-zb}$&$a_{\rm zb}$&$a_{\rm w}$&$c_{\rm w}$&
$\frac{c}{a}$&u\\
\hline
{\bf SiC}&HF&+0.0004&4.3793&3.0914&5.0728&1.6409&0.3763\\
&HF+corr&+0.0007&4.3862&&&&\\
&LDA\cite{karch94}&+0.0007&4.3445&3.0692&5.0335&1.64&ideal\\
&LDA\cite{kaeckell94}&+0.00007&4.2907&3.020&5.012&1.66&0.3758\\
&exp.&&4.3596&3.0763&5.0480&1.6409&\\
\hline
{\bf AlN}&HF&-0.0030&4.3742&3.1002&4.9888&1.6092&0.3805\\
&HF+corr&-0.0036&&&&&\\
&LDA\cite{yeh92}&-0.0014&4.365&3.099&4.997&1.612&0.381\\
&LDA\cite{miwa93}&&4.421&3.144&5.046&1.605&0.381\\
&LDA\cite{kim96}&&4.32&3.06&4.91&1.60(5)&0.383\\
&exp.&&&3.110\dots 3.1127&4.9798\dots 4.982&1.5998\dots 1.6019&\\
\hline
{\bf GaN}&HF&-0.0010&4.5215&3.2011&5.1970&1.6235&0.3775\\
&HF+corr&-0.0013&&&&&\\
&HF\cite{pandey96}&&&3.199&5.176&1.618&0.38\\
&LDA\cite{yeh92}&-0.0007&4.364&3.095&5.000&1.633&0.378\\
&LDA\cite{miwa93}&&4.446&3.146&5.115&1.626&0.377\\
&LDA\cite{kim96}&&4.46&3.17&5.13&1.62&0.379\\
&exp.&&&3.160\dots 3.190&5.125\dots 5.190&1.6249\dots 1.6279&\\
\hline
{\bf InN}&HF&-0.0015&4.9870&3.5428&5.7287&1.6170&0.3784\\
&HF+corr&-0.0023&&&&&\\
&LDA\cite{yeh92}&-0.0008&4.983&3.536&5.709&1.615&0.380\\
&LDA\cite{kim96}&&4.92&3.53&5.54&1.57&0.388\\
&exp.&&&3.5446&5.7034&1.6090&\\
\end{tabular}
\end{table}

\begin{table}
\caption{\label{coh}Cohesive energies per unit cell
in Hartree; their ratio to corresponding
experimental values is given in parentheses; for
comparison, literature  data from LDA calculations are also
reported.  } 
\begin{tabular}{c||c|c|c|c}
&SiC&AlN&GaN&InN\\
\hline
$E_{\rm coh}^{\rm HF}$&-0.329(70\%)&-0.287(67\%)&-0.185(56\%)&-0.132(45\%)\\
\hline
$E_{\rm coh}^{\rm HF+corr}$(Basis A)&-0.418(89\%)&-0.380(89\%)&
-0.288(86\%)&-0.236(81\%)\\
\hline
$E_{\rm coh}^{\rm HF+corr}$(Basis B)&-0.440(93\%)&-0.400(93\%)&-0.307
(92\%)&-0.257(88\%)\\
\hline
$E_{\rm coh}^{\rm LDA}$&-0.622\protect{\cite{kaeckell94}}(132\%)&---&-0.397
\protect{\cite{kim94}}(119\%)&---\\
&&&-0.478\protect{\cite{jhi95}}(144\%)&\\
\hline
$E_{\rm coh}^{\rm exp}$&-0.471&-0.429&-0.333&-0.293
\end{tabular}
\end{table}

\begin{table}
\caption{\label{bulk}Bulk moduli in Mbar for 
the zinc-blende structure; deviations from the 
average experimental values (Ref. \protect{\onlinecite{kaeckell94}} and
\protect{\onlinecite{kim96}} and references therein) are given in parentheses;
for comparison, literature  data from LDA calculations are also
reported. } 
\begin{tabular}{cc||c|c|c|c}
&&SiC&AlN&GaN&InN\\
\hline
$B_{\rm HF}$&zb&2.54(+13\%)&2.18(+10\%)&2.54(+17\%)&1.59(+27\%)\\
$B_{\rm HF+corr}$&zb&2.44(+9\%)&&&\\
\hline
$B_{\rm LDA}$&zb&2.22\protect{\cite{kaeckell94}}&2.03\protect{\cite{kim96}}&
2.01\protect{\cite{kim96}}&1.39\protect{\cite{kim96}}\\
&w&2.10\protect{\cite{kaeckell94}}&2.02\protect{\cite{kim96}}&
2.07\protect{\cite{kim96}}&1.46\protect{\cite{kim96}}\\
\hline
$B_{\rm exp}$&zb&2.24&---&---&---\\
&w&2.23&1.85-2.12&1.88-2.45&125
\end{tabular}
\end{table}

\newpage

\newpage


\begin{thebibliography}{10}

\bibitem{davis93} R.~F. Davis, Physica B {\bf 185}, 1 (1993).

\bibitem{yeh92} C.-Y. Yeh, Z.W. Lu, S. Froyen, and A. Zunger, Phys. Rev. B
{\bf 46}, 10086 (1992).

\bibitem{min92} B.~J. Min, C.~T. Chan, and K.~M. Ho, Phys. Rev. B
{\bf 45}, 1159 (1992).
 
\bibitem{miwa93} K. Miwa, and A. Fukumoto, Phys. Rev. B {\bf 48}, 7897
(1993).

\bibitem{kim96} K. Kim, W.~R.~L. Lambrecht, and B. Segall, Phys. Rev. B
{\bf 54}, Sep. (1996).

\bibitem{karch94} K. Karch, P. Pavone, W.Windl, O. Sch{\"u}tt, and
D. Strauch, Phys. Rev. B {\bf 50}, 17054 (1994).

\bibitem{kaeckell94} P. K{\"a}ckell, B. Wenzien, and F. Bechstedt,
Phys. Rev. B {\bf 50}, 17037 (1994).

\bibitem{pisani88}C. Pisani, R. Dovesi, and C. Roetti, in
{\it Hartree-Fock Ab initio Treatment of Crystalline Systems},
edited by G. Berthier {\it et al.},
Lecture Notes in
Chemistry, Vol.\ {\bf 48}
(Springer, Berlin, 1988)


\bibitem{crystal92} R. Dovesi, V.~R. Saunders, and C. Roetti, {\em
Computer Code Crystal92}, Gruppo
  di Chimica teorica, University of Torino and United Kingdom Science
  and Engineering Research Council Laboratory, Daresbury, 1992.

\bibitem{stoll92} H. Stoll,  Phys. Rev. B {\bf 46}, 6700 (1992).


\bibitem{bergner93} A. Bergner, M. Dolg, W. K\"{u}chle, H. Stoll and
H. Preuss, Mol. Phys. {\bf 80},  1431  (1993).

\bibitem{schwerdt95} P. Schwerdtfeger, Th. Fischer, M. Dolg, G.
Igel-Mann, A. Nicklass, H. Stoll and A. Haaland, J. Chem. Phys. {\bf
102}, 2050 (1995); T. Leininger, A. Nicklass, H. Stoll, M. Dolg and
P. Schwerdtfeger, J. Chem. Phys. {\bf 105}, 1052 (1996).

\bibitem{13vepp} M. Dolg, unpublished results; T. Leininger, in
preparation.

\bibitem{molpro94} {\sc Molpro} is a package of {\em ab initio}
programs written by H.-J. Werner and P.J. Knowles, with contributions
from J.  Alml\"of, R.D. Amos, A. Berning, C. Hampel, R. Lindh, W.
Meyer, A. Nicklass, P. Palmieri, K.A. Peterson, R.M. Pitzer, H.
Stoll, A.J. Stone, and P.R. Taylor; C. Hampel, K. Peterson and H.-J.
Werner, Chem. Phys. Lett. {\bf 190}, 1 (1992); P.J. Knowles, C.
Hampel and H.-J. Werner, J. Chem. Phys. {\bf 99}, 5219 (1993).

\bibitem{pandey96} R. Pandey, M. Causa, N. M. Harrison and M. Seel,
J. Phys.: Condens. Matter {\bf 8}, 3993 (1996). 

\bibitem{paulus95} B. Paulus, P. Fulde, and H. Stoll, Phys. Rev. B
{\bf 51},  10572  (1995).

\bibitem{paulus96} B. Paulus, P. Fulde, and H. Stoll, Phys. Rev. B
{\bf 54}, 2556 (1996).

\bibitem{foster60} J.M. Foster and S.F. Boys, Rev. Mod. Phys. {\bf
32},  296  (1960).

\bibitem{dunning89} T.~H. Dunning, J. Chem. Phys. {\bf 90},  1007
(1989).

\bibitem{lei92} T. Lei, T.~D. Moustakas, R.~J. Graham, Y. He,
and S.~J. Berkowitz, J. Appl. Phys. {\bf 91}, 4933 (1992).

\bibitem{strite92} S. Strite, J. Ruan, D.~J. Smith J. Sariel, N. Manning, 
H. Chen, W.~J. Choyke, and H. Morko\c{c}, Bull. Am. Phys. Soc. {\bf 37},
346 (1992)

\bibitem{landoltiii17a} Landolt/B\"{o}rnstein, {\em Physics of group
IV elements and III-V compounds} (Springer, Berlin, 1982), Vol.~III 17a.

\bibitem{kalvoda97} S. Kalvoda, B. Paulus, P. Fulde, and H. Stoll,
Phys. Rev. B (1997), in press.

\bibitem{harrison89} W.~A. Harrison, {\it Electronic structure and the
properties of solids}, Dover Publications, Inc., New York (1989).

\bibitem{farid91} B. Farid and R. Godby, Phys. Rev. B {\bf 43},
14248  (1991).

\bibitem{moore} C.E. Moore, Atomic Energy Levels, Circ. Nat. Bur.
Stand. 467 (US Dept. of Commerce, Washington DC, 1958)
        
\bibitem{kim94} K. Kim, W.~R.~L. Lambrecht, and B. Segall, Phys. Rev. B
{\bf 50}, 1502 (1994).

\bibitem{jhi95} S.-H. Jhi and J. Ihm, Phys. Stat. sol. (b) {\bf 191},
387 (1995).



\end{thebibliography}
\end{document}